\documentclass[twocolumn,aps,prd,amsmath,amssymb,floatfix,amsmath,superscriptaddress]{revtex4-2}

\usepackage{graphicx}
\usepackage{dcolumn}
\usepackage{bm}
\usepackage{braket}
\usepackage{color,epsfig}
\usepackage{bm}
\usepackage{slashed}
\usepackage{hyperref}
\usepackage{subfigure}
\usepackage{feynmf}
\usepackage{calrsfs}

\usepackage{booktabs}
\usepackage{siunitx}

\newcommand{\bea}{\begin{eqnarray}}
\newcommand{\eea}{\end{eqnarray}}
\newcommand{\be}{\begin{equation}}
\newcommand{\ee}{\end{equation}}

\renewcommand\vec{\bm}

\begin{document}


\title{Graviton Mass Bounds in Very Special Relativity\\
from Binary Pulsar's Gravitational Waves}

\author{Alessandro Santoni}
\email{asantoni@uc.cl}
\affiliation{Institut f\"ur Theoretische Physik and Atominstitut,
 Technische Universit\"at Wien,
 Wiedner Hauptstrasse 8--10,
 A-1040 Vienna, Austria}
\affiliation{Instituto de Física, Pontificia Universidad de Católica de Chile, \\
 Avenida Vicuña Mackenna 4860, Santiago, Chile }

\author{Jorge Alfaro}
\email{jalfaro@fis.puc.cl}
\affiliation{Instituto de Física, Pontificia Universidad de Católica de Chile, \\
 Avenida Vicuña Mackenna 4860, Santiago, Chile }

\author{Alex Soto}
\email{alex.soto@newcastle.ac.uk}
\affiliation{School of Mathematics, Statistics and Physics, Newcastle University,\\ 
Newcastle upon Tyne, NE1 7RU, UK}


\date{\today}

\begin{abstract}
In this work we study the gravitational radiation produced by a keplerian binary system within the context of Very Special Linear Gravity (VSLG), a novel theory of linearized gravity in the framework of Very Special Relativity (VSR) allowing for a gauge-invariant mass $m_g$ of the graviton. For this task, we exploit Effective Field Theory's techniques, which require, among others, the calculation of the squared amplitude of the emission process and therefore the polarization sum for VSLG gravitons. Working in the radiation zone and using the standard energy momentum tensor's expression for keplerian binaries, we derive and study the properties of the VSLG energy loss and period decrease rates, also verifying they reduce to the correct General Relativity limit when sending $m_g\to0$. Finally, using astronomical data from the Hulse-Taylor binary and the Double Pulsar J0737-3039, we obtain an upperbound on the VSLG graviton mass of $m_g\sim 10^{-21}eV$ that, while being comparable to bounds obtained in this same way for other massive gravity models, is still weaker than the kinematical bound $\sim 10^{-22}eV$ obtained from the combined observation of the astronomical events GW170817 and GRB170817A, which should still hold in VSLG.
\end{abstract}

\maketitle


\section{Introduction}
The observation of gravitational waves (GW) is one of the most notable successes of General Relativity (GR). Binary stars played an important role in such progress: indirect measurements from binary pulsars in the seventies \cite{1982ApJ...253..908T,Weisberg:1984zz} served as initial evidence for this prediction. The definitive confirmation was the first direct detection of gravitational waves from a black hole merger by the LIGO/VIRGO collaboration \cite{LIGOScientific:2016aoc} in 2015. The unique precision of this measure opens interesting possibilities to test modifications of GR, imposing strong constraints on such models. Nevertheless, binary stars still offers a fantastic playground to test for many predictions of GR and, in general, fundamental physics \cite{kramer2016pulsars, shao2020new}.\\

Extensions or alternatives to GR have been widely studied for years. Despite the success of GR, open problems such as Dark Energy, the cosmological constant or tensions in cosmology have suggested the possibility of modifications to the Einstein's picture to give an explanation to such questions, while still containing GR as some suitable limit. Different studies on these alternative models have been carried out considering binary systems. For example, the rate of energy loss and orbital period decay was recently analysed in $f(R)$ models \cite{Narang:2022jkv}. Similarly, gravitational radiation was studied for certain massive gravity models \cite{Poddar:2021yjd}.\\
Massive gravity theories have a long story. The pioneering work of Fierz and Pauli presented for the first time a lagrangian for a massive spin-2 particle \cite{Fierz:1939ix}. However, it was discovered later that a discontinuity with GR appears in the massless limit of the theory, the so-called vDVZ discontinuity \cite{vanDam:1970vg,Zakharov:1970cc}. Also, massive gravitons possess five degrees of freedom (d.o.f) instead of two as in the massless case. This fact has consequences in the manifestation of ghost modes \cite{Boulware:1972yco}. Solutions to those problems, like the Vainshtein mechanism \cite{Vainshtein:1972sx} for example, have been already studied and implemented in models such as DGP \cite{Dvali:2000hr,Dvali:2000rv} and dRGT \cite{deRham:2010kj} or New Massive Gravity (NMG) in three-dimensions \cite{Bergshoeff:2009hq}. For the interested reader, we refer to the review \cite{deRham:2014zqa} and references therein for more details on these theories.\\

The introduction of a graviton mass in the models above breaks GR gauge symmetry. However, an interesting possibility to add masses without breaking gauge invariance can be found in the Very Special Relativity (VSR) framework. VSR appeared as a novel way to explain neutrino masses without the addition of new particles \cite{vsr2}, just by considering as the symmetry group of nature some special Lorentz's subgroup, like $SIM(2)$, rather than the Lorentz group \cite{vsr1}, introducing in this way a preferred lightlike spacetime direction labelled by the four-vector $n^\mu$. \\
Interestingly, the VSR symmetry group can always be enlarged to the full Lorentz group by the addition of discrete transformations like P, T or CP, recovering in such manner the Special Relativity (SR) picture. Thus, VSR effects should be absent in contexts where these discrete symmetries are present. This feature becomes attractive in the gravitational sector and cosmology in light of the Sakharov conditions for baryogenesis \cite{Sakharov:1967dj}, as a result of which CP violation is needed to account for the observed baryon-asymmetry of the universe.\\

Apart from neutrino masses, it has been shown that, in VSR, gauge boson masses can be added in a gauge-invariant way. This issue has been explored in general for any non-abelian field in \cite{Alfaro:2013uva}, and in particular within the electroweak model \cite{Alfaro:2015fha} and for the photon \cite{qed1,qed2}. Recently, it has been presented in \cite{grav3} a way to describe a VSR massive graviton model, while preserving gauge invariance and the usual number of d.o.f, as it happens in the other VSR extensions. \\
Therefore, this model, that we call Very Special Linear Gravity (VSLG), does not present problems with ghosts as it happens in other massive graviton realisations.\\

Other forms of gravity having as starting point VSR have been studied using algebra deformations, analysing line elements invariants under $DISIM_b(2)$ \cite{grav1,grav2}, and showing an interesting connection with Finsler geometries. These approaches are different from the one in \cite{grav3} since, in the latter, gravity is studied as a linear perturbation of the Minkowski metric $\eta = diag\{+,-,-,-\}$ with the addition of new lagrangian terms invariant under the $SIM(2)$ group. Note that VSLG also fundamentally differs from any spurion effective field theory extension of linearized gravity, like \cite{kostelecky2018lorentz}, since the $SIM(2)$ algebra does not allow for new invariant background vectors or tensors, leading to the non-locality of the VSLG additional Lagrangian terms. \\

Thus, the aim of this work is to study binary systems and their gravitational emission in the context of VSR, focusing on the VSLG formulation developed in \cite{grav3}.\\
The outline of the present paper is as follows: in section II we present the VSLG lagrangian. We show the non-relativistic gravitational potential for the model and we present the derivation of the rate of gravitational energy loss in an emission process under an effective field theory perspective. In section III, we discuss the energy momentum tensor used for the binary system in the emission process, considering the regime of validity of our approximations. Section IV is devoted to show the way to compute the angular integrals appearing in the expression for the energy loss rate. In Section V, we put together all the previous computations to obtain the VSR rate of orbital period decrease, discussing some of its novel features. In section VI, we present indicative bounds for the VSR graviton mass considering the previous results combined with experimental data from binaries and, finally, we summarise and conclude our work in section VII. \\To avoid clutter with the long expressions of VSLG, we have left in the appendices the complete expressions of some relevant quantities, which are referenced in the main text when necessary, together with a few more details on their computation.\\

\section{Very Special Linear Gravity}
In \cite{grav3}, it was shown that the most general quadratic $SIM(2)-$invariant lagrangian for a graviton $h_{\mu \nu}$, containing up to two derivatives, is
\begin{equation}
\label{hlag}
\mathcal{L}_h = \frac{1}{2} h^{\mu \nu} \mathcal{O}_{\mu \nu \alpha \beta}
h^{\alpha \beta} \, ,
\end{equation}
where the operator $\mathcal{O}$ is fully written in appendix \ref{appLagandFeyn}. This operator contains terms involving $N^{\mu} \equiv \frac{n^{\mu}}{n\cdot \partial}$, with $n^\mu = (n^0,\vec n)$ being the usual lightlike vector characterizing $SIM(2)$ models \cite{vsr1,vsr2}. Despite these extra terms, it was also shown in \cite{grav3} that the physical degrees of freedom of this model are still only two as in GR. \\
Now, we are interested in seeing the novel effects generated by VSLG when interacting weakly with matter. For this purpose, we will couple the graviton to matter in the usual gauge-invariant and linear way \cite{Donoghue:2017pgk}
\begin{equation}
\label{lagmatt}
\mathcal{L}_{int} = - \frac{\kappa}{2} h_{\mu \nu}T^{\mu \nu} \, , \,\,\, \kappa \equiv \sqrt{ 32 \pi G } \, ,
\end{equation}
with $T^{\mu \nu}$ representing energy-momentum tensor (EMT) of matter. In principle, $T^{\mu \nu}$ could reflect any kind of matter that we could be interested in. \\
In the next sections, we will use a keplerian binary system as a source. However, to get familiar with calculations in VSLG, we decide to start with the computation of its gravitational potential. For this purpose, we consider the case of a massive scalar field $\phi$ and its scattering through the exchange of a graviton, as shown below in Fig.\ref{fig:diagram-tree}. 

\begin{figure}[h!!]
    \includegraphics[width=3.9cm]{"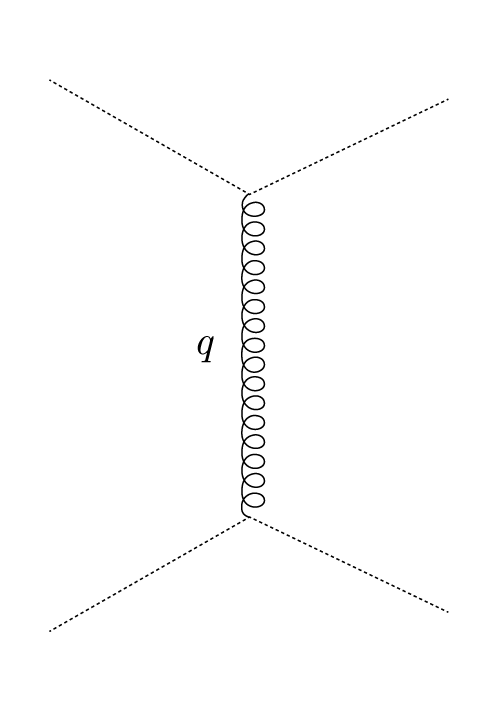"}
    \caption{Tree level diagram for the scattering of two scalar fields through the exchange of a graviton of momentum $q$.}

    \label{fig:diagram-tree}
\end{figure}

\subsection{Non Relativistic Gravitational Potential}
For a scalar field we have
\begin{equation}
\label{emtscalar}
T_{\mu \nu}=\partial_\mu \phi \, \partial_\nu \phi-\frac{1}{2}\eta_{\mu \nu}(\partial_\rho \phi \, \partial^{\rho} \phi - m^2 \phi^2) \, .
\end{equation}
To calculate the amplitude $\mathcal{M}$ for a scattering of two scalar particles mediated by a graviton at tree level, we will need the graviton propagator. In order to obtain it, we have to add the following gauge fixing term to the VSLG lagrangian
\begin{equation}
\label{gaugefixing}
\mathcal{L}_{GF}=\xi \partial_{\mu} \left( h^{\mu \nu} - \frac{1}{2}
\eta^{\mu \nu} h \right) \partial^{\lambda} \left( h_{\lambda \nu} -
\frac{1}{2} \eta_{\lambda \nu} h \right) .
\end{equation}
The relevant expressions of the Feynman rules for the graviton propagator and the vertex $h\phi \phi$ are left in the appendix \ref{appLagandFeyn}. The gravitational potential in the non-relativistic limit can be obtained easily from the scattering amplitude $\mathcal{M}$, and it is given by
\begin{equation}
\label{vsrpotentialq}
  V (\vec{q}) = - \frac{4 \pi G m^2}{| \vec{q} |^2 + m^2_g} \left( 1
  + \frac{m^2_g}{(\hat{n} \cdot \vec{q})^2} \right)^2 ,
\end{equation}
being $\hat n $ the unitary VSR spatial vector.
We can see that, as expected, the $ \hat  n-$independent term coincides with the expression for a Yukawa potential. However, the terms containing $\hat n \cdot \vec{q}$ in the denominator are infrared divergent, therefore, they must be regularised. In fact, these divergences would imply the existence of infinite forces for certain angles, leading to an ill-defined physical situation. The existence of such divergences is a quite common feature of VSR models. In relativistic cases, the Mandelstam-Leibbrandt prescription has been used to regulate them \cite{Alfaro:2017umk}. Nevertheless, this trick is not helpful in the non-relativistic case. Recently, a new prescription to manage VSR divergences also in the non-relativistic case was presented in \cite{Alfaro:2023qib}. The crucial point is that, with this recipe, the $SIM(2)$ limit of integrals like
\begin{equation}
\int \frac{d^3 q}{(2\pi)^3} \frac{e^{-i\vec{q}\cdot \vec{x}}}{| \vec{q} |^2 + m^2_g}\frac{1}{(\vec n \cdot \vec{q})^l}
\end{equation}
vanishes. Hence, following this technique, the position-space expression for the non-relativistic gravitational potential in VSLG would be a standard Yukawa potential
\begin{equation}
\label{vsrpotential}
V (r) = - \frac{G m^2}{r}e^{- m_g r} .
\end{equation}
Notice that the gravitational potential for this case is independent of the direction of $\hat n$, as typical in many other massive graviton's models. However, the main difference is that here the physical degrees of freedom are still two, as in the standard GR case.

\subsection{Gravitational Energy Loss Rate}

\begin{figure}[h!!]
    \includegraphics[width=4cm]{"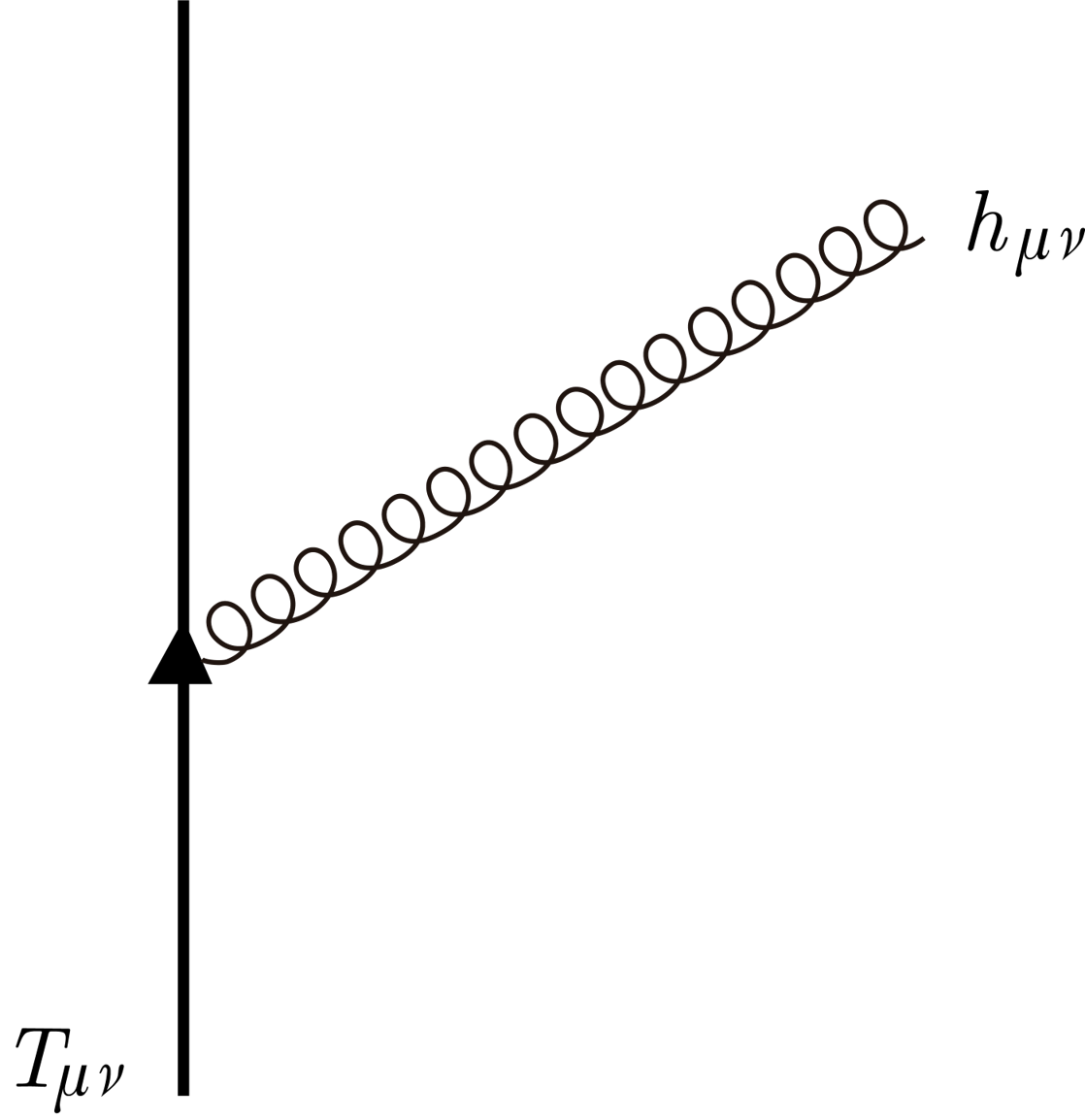"}
    \caption{Diagramatic representation of the emission of a graviton, derived from the gauge invariant linear coupling $\frac{\kappa}{2} T_{\mu \nu}(x) \, h^{\mu \nu} (x)$ of the field $h_{\mu\nu}$ with the classical source $T$.}
    \label{fig:diagram-emission}
\end{figure}

Now, we turn our attention to the main goal of the paper: the calculation of the rate of energy lost by gravitational radiation for a binary system. For this purpose, we will stick to the effective field theoretical approach \cite{goldberger2007effective,goldberger2010gravitational,goldberger2022effective,sturani2021fundamental}. In particular, inspired by \cite{Poddar:2021yjd}, we are thinking of a tree level emission process like the one shown in Fig.\ref{fig:diagram-emission}, where $T_{\mu\nu}$ represents the binary system and $h_{\mu \nu}$ the emitted gravitational wave.\\
\\
The tree-level amplitude of such a process would be
\begin{equation}
    {A}_\lambda = - \frac{\kappa}{2} \tilde T_{\mu \nu}(k) \, \epsilon_\lambda^{* \;\mu \nu} (k) \, ,
\end{equation}
with $\tilde T_{\mu\nu}$ being the EMT expression in momentum space, $\epsilon_\lambda^{\mu\nu}$ being the graviton's polarization and $\lambda=1,2$ labelling the different physical polarizations of VSLG. Note that, since $T_{\mu\nu}(x)$ is real for our case of interest, then we have the relation $\tilde T_{\mu\nu}(-k)=\tilde T_{\mu\nu}^{\,*}(k)$.\\
From the unpolarized squared amplitude we can easily find the infinitesimal emission probability $d \Sigma$ \cite{goldberger2007effective}
\begin{eqnarray} \label{emrate1}
    d\Sigma &=& \sum_\lambda |A_\lambda|^2 \frac{d^3 k}{(2 \pi)^3 2 \omega}  \\
    &=& \frac{\kappa^2}{ 8 (2 \pi )^3} \sum_\lambda |\tilde T_{\mu \nu}(k) \, \epsilon_\lambda^{* \;\mu \nu} (k)|^2 \frac{d^3k}{\omega}\,. \nonumber
\end{eqnarray}
Defining the polarization sum $S^{\mu \nu \alpha \beta}$ as
\begin{eqnarray}
    S^{\mu \nu \alpha \beta} (k) = \sum_\lambda \epsilon_\lambda^{\mu \nu}(k) \epsilon_\lambda^{* \; \alpha \beta} (k) \,,
\end{eqnarray}
we can rewrite Eq.~\eqref{emrate1} as
\begin{equation} \label{emrate2}
    d\Sigma = \frac{\kappa^2}{ 8 (2 \pi )^3} \tilde T^{\,*}_{\mu \nu}(k) \tilde T_{\alpha \beta}(k) S^{\mu\nu\alpha\beta}(k) \frac{d^3k}{\omega} \, ,
\end{equation}
and, being $\omega^2= |\vec k|^2+ m_g^2$ for on-shell physical graviton modes in VSLG, we can replace 
\begin{equation}
    d^3k = |\vec k|^2 d|\vec k| d\Omega_k = \omega^2 \sqrt{1- \frac{m_g^2}{\omega^2} } d\omega d \Omega_k \, ,
\end{equation}
in Eq.~\eqref{emrate2}, obtaining
\begin{equation} 
    d\Sigma = \frac{\kappa^2}{ 8 (2 \pi )^3} \tilde T^{\,*}_{\mu \nu}(k) \tilde T_{\alpha \beta}(k) S^{\mu\nu\alpha\beta}(k) \rho(\omega) \, \omega \, d\omega \, d \Omega_k   \, ,
\end{equation}
where we defined $\rho(\omega) \equiv \sqrt{1- \frac{m_g^2}{\omega^2}}$. \\

\noindent
To derive the differential rate of radiated energy is sufficient to multiply $d\Sigma$ by the energy $\omega$ of the emitted GW and divide it by the observation time $T$. Thus, integrating over all emission angles and frequencies, we get the total energy loss rate $\dot E \equiv \frac{dE}{dt}$
\begin{eqnarray} \label{dedt1}
    \dot E &=& \int \frac{\omega}{T} \, d\Sigma = \\
    &=& \frac{\kappa^2}{ 8 (2 \pi )^3 T} \, \int  \rho(\omega) \, \omega^2 \tilde T^{\,*}_{\mu \nu} \tilde  T_{\alpha \beta} S^{\mu\nu\alpha\beta} \, d\omega \, d \Omega_k \,. \nonumber
\end{eqnarray}
The explicit expression for $S^{\mu\nu\alpha\beta}$, that we figured out for this paper, is included in the Appendix \ref{appPolar}. In particular, the relevant formula for the calculations in this work is given in Eq.~\eqref{polsum}.\\
At this point, it is convenient to re-express the $\tilde T \tilde T S $ spacetime contraction just as a spatial contraction 
\begin{equation}
    \tilde  T^{\,*}_{\mu\nu} \tilde  T_{\alpha \beta} S^{\mu\nu\alpha\beta} \to
    \sum_{i j k l} \tilde  T^{\,*\,ij} \, \tilde  T^{\, kl} \Lambda^{ijkl} \, ,
\end{equation}
therefore transforming Eq.~\eqref{dedt1} into (repeated spatial indices are summed on)
\begin{equation} \label{dedt2}
    \dot E = \frac{\kappa^2}{ 8 (2 \pi )^3 T} \,  \int  \rho(\omega) \, \omega^2d\omega \int \tilde T^{\,*\,ij} \, \tilde T^{\, kl} \Lambda^{ijkl} \,  \, d \Omega_k  \,.
\end{equation}
To do that we exploit the energy momentum conservation
\begin{equation}
    k_\mu \tilde T^{\mu \nu} = 0 \rightarrow \omega \tilde T^{0 \nu } = k^i \tilde T^{i \nu} \,,
\end{equation}
as well as the relations
\begin{equation} \label{timetospatial}
     \left\{\begin{array}{l} 
    \tilde T^{0i} = \frac{k^j}{\omega} \tilde T^{ij} \,,\\
    \tilde T^{00} = \frac{k^i}{\omega} \tilde T^{0i} = \frac{k^i k^j}{\omega^2} \tilde T^{ij}\,.
    \end{array} \right.  
\end{equation}
Let's also introduce the unitary vector $\hat k^i$
\begin{equation}
    \hat k^i \equiv \frac{ k^i}{\omega  \sqrt{1 -\frac{m^2_g}{\omega^2}}} = \frac{k^i}{\omega \rho (\omega)} \, ,
\end{equation}
so that
\begin{equation}
    N^\mu \equiv \frac{n^\mu }{n\cdot k} = \frac{n^\mu}{n^0 \omega \left (1- \rho  \frac{\vec n}{n^0} \cdot \hat k \right )} =\frac{n^\mu /n^0}{ \omega \, R(\omega,\hat k ,n^\mu)} \, ,
\end{equation}
with $R(\omega, \hat k,n^\mu) \equiv 1- \rho \,\frac{\vec n}{n^0} \cdot \hat k $ and $n^\mu$ labelling the VSR preferred spacetime direction. From hereon, thanks to the $n-$rescaling symmetry of VSR, we will fix $n^0=1$, implying $\vec n \to \hat n$. \\
Note that $\Lambda ^{ijkl}$ must be symmetric under the interchange of $i j \Longleftrightarrow k l$, due to the symmetry properties of $S^{\mu\nu\alpha\beta}$ and since it is contracted with $\tilde  T^{*\,ij} \tilde  T^{\, kl}$, as also stressed in Appendix \ref{appLamb}, where we additionally include some other useful details for its computation. \\
Going carefully through all the calculations, we finally obtain the expression \eqref{TTlambda} for $\Lambda^{i j k l}(\omega,\hat k ,\hat n)$.\\

\section{Energy momentum tensor}

In this section we will discuss the expression of the EMT appearing in our calculation. Due to the linear nature of VSLG and the huge distance from which we observe astrophysical phenomena, the only relevant component that enters in $T^{\mu\nu}$ is the one from the classical source \cite{Poddar:2021yjd}, which in our case describes an inspiraling two-body system.\\
Working in the non-relativistic or low-velocity limit, we can make use of the known EMT's formula for a keplerian binary of total mass $M=m_1+m_2$
\begin{equation} \label{Tdef}
    T^{\mu\nu} (t,\vec x) = \mu \, U^\mu U^\nu \delta^3(\vec x - {\vec r}(t)) \, ,
\end{equation}
where $ {\vec r}(t)$ and $U^\mu = (1, \dot r_x , \dot r_y ,0)$ are the trajectory and non-relativistic four-velocity in the $x-y$ orbital plane of the reduced mass $\mu= m_1 m_2 / M $. Defining $b$ and $e$ respectively as the semi-major axis and the eccentricity of the keplerian orbit, we can parametrize the motion in function of the eccentric anomaly $\phi= \phi(t)$ \cite{celmec} as follows
\begin{eqnarray}
    &&\vec r(t) = b \left ( \cos \phi-e , \sqrt{1-e^2} \sin\phi, 0 \right) \,,\nonumber \\
    && \Omega t = \phi - e \sin\phi \,\,\,\text{with} \,\,\, \Omega = \sqrt{ \frac{G M}{b^3}} \,,
\end{eqnarray}
with $\Omega$ representing the fundalmental frequency defined by the revolution period $P_b \to \Omega = 2\pi/P_b$.\\
At this point let's observe that, since the EMTs in Eq.~\eqref{dedt2} are the momentum space version, we can write them as
\begin{eqnarray}
    \tilde T^{ij} (\omega ,\vec k) &=& \int dt \int d^3x \, e^{i(\omega t - \vec k \cdot \vec x)} T^{ij}(t,\vec x) \nonumber \\
    &=& \int d^3 x \,e^{ i \vec k \cdot \vec x} \, \tilde T^{ij}(\omega, \vec x) \,.
\end{eqnarray}
Defining by $d$ the distance between us and the source, we will work as usual in the so-called ‘‘far zone" or ‘‘radiation zone", which implies the following hierarchy of lenghts $ b<< \lambda<< d $, with $\lambda$ being the wavelength of the emitted radiation. Thus, being $\vec k \cdot \vec x \sim  \frac{1}{\lambda} \, b<<1$, we can approximate $e^{i \vec k' \cdot \vec x} \sim 1$ and write
\begin{eqnarray}
   \tilde  T^{ij} (\omega ,\vec k) \simeq \int d^3 x \,\tilde  T^{ij}(\omega, \vec x) \equiv \tilde  T^{ij}(\omega)\,.
\end{eqnarray}
Note that this approximation would also imply an upper limit in the $\omega $-integral. In fact, being $v<<1$ the typical velocity of the system (where we are working with the velocity of light $c=1$), we have
\begin{equation}
    k << \frac{1}{b} \sim \frac{1}{v}\Omega \to \omega << \sqrt{m_g^2 +\frac{1}{v^2}\Omega^2 } \,,
\end{equation} 
Then, considering systems for which $ m_g \lesssim \Omega$, as in the cases considered in the rest of the paper, we would have the upperlimit
\begin{equation}\label{omegaupper}
    \omega << \frac{1}{v}\Omega \, .
\end{equation}
However, we will not worry about this constraint, because at the end it is already taken care of by the fast convergence of the integrand, as pointed out later in Section \ref{sec:mgbound}.\\
Furthermore, let's observe the following: from the conservation equation $\partial_\mu T^{\mu\nu} (t,\vec x) = 0$, contracting it with another $\partial_\nu$, we can find that $ \partial_i \partial _j T^{ij} (t,\vec x) = \partial_0^2 T^{00} (t,\vec x)$ or equivalently
\begin{equation}
     \partial_i \partial _j \tilde T^{ij} (\omega ,\vec x) = - \omega^2 \, \tilde T^{00} (\omega ,\vec x) \, ,
\end{equation}
that multiplied by $x^k x^l$ and integrated over $d^3x$ becomes, using integration by parts
\begin{eqnarray}\label{Tomega}
   \tilde  T^{kl} (\omega) &=& -\frac{\omega^2}{2} \int dt \, e^{i\omega t} \int d^3 x\, T^{00} (t,\vec x)  x^k x^l \\
    &=& -\frac{\omega^2}{2} \int dt \, e^{i\omega t} Q^{kl}(t) = -\frac{\omega^2}{2} \tilde Q^{kl}(\omega) \,, \nonumber
\end{eqnarray}
from which we see, as expected, that the first contribution in our expansion comes from the quadrupole moment $Q^{ij}$ of the source $\sim \int d^3x \, T^{00} x^k x^l$.\\ 
Using the definition \eqref{Tdef} in the expression for $Q^{ij}$, we get
\begin{eqnarray} \label{qtspace}
    Q^{kl} (t) &=& \mu \int d^3 x \, \delta^3(\vec x -\vec r(t)) x^k x^l \nonumber \\
    &=& \mu \, r^k (t) \, r^l(t) \,.
\end{eqnarray}
Replacing the expression \eqref{Tomega} for $T^{kl} (\omega)$ back into the energy loss rate and using the definition of $\kappa$, we finally end up with
\begin{equation} \label{dedt3}
    \dot E = \frac{ G}{ 8 \pi ^2 T} \, \int d\omega \rho (\omega) \,  \omega^6 \, \tilde Q^{\,*\,ij} \, \tilde Q^{\, kl} \left [\int \, \Lambda^{ijkl} \, d\Omega_k \right ] \,.
\end{equation}

\subsection{System Periodicity and Fourier}

At this point, we can further simplify our calculations by considering the (almost) periodic nature of binaries, that is justified by the smallness of the rate of period change due to gravitational emission in the initial phase of the binary inspiral (see for example 
Table \ref{tab:pulsarsdata}). In fact, due to the periodicity, we can express the quadrupolar moment as a Fourier series
\begin{equation}
    Q^{ij}(t)= \sum_N Q_N^{ij} e^{i \omega_N t} \, ,
\end{equation}
with $\omega_N = N\Omega$. Then, its fourier transform will be
\begin{equation}
    \tilde Q^{ij}(\omega) = 2 \pi \sum_N \delta (\omega- \omega_N) \, Q_N^{ij}\, ,
\end{equation}
where we remember the standard definition of the Fourier coefficients
\begin{equation}\label{qfourier}
    Q_N^{ij} = \frac{1}{P_b} \int_0 ^{P_b} dt \, e^{- i \omega_N t} Q^{ij}(t) \, .
\end{equation}
Replacing this expression in the energy loss rate, we can integrate over the frequency $\omega$ to remove the deltas and set $\omega = \omega_N$, since
\begin{equation}
    \int _{m_g}^\infty d\omega \, \delta(\omega-\omega_N)\delta(\omega-\omega_M) \to \delta(\omega_N-\omega_M) = \frac{1}{\Omega} \delta_{NM} \,.
\end{equation}
Then, using the $\delta_{NM}$ to remove one of the sums that derives from the two $\tilde Q$ and taking $T=P_b$, we obtain
\begin{equation} \label{dedt4}
    \dot E = \frac{ G}{ 4 \pi } \sum_{N_{min}} \, \rho (\omega_N) \,  \omega_N^6 \,  Q^{\,*\,ij}_N \, Q^{\, kl}_N \left [\int \, \Lambda^{ijkl} \, d\Omega_k \right ] \,,
\end{equation}
whit $N_{min} = \lceil \frac{m_g}{\Omega} \rceil$, where the brackets $\lceil \, \rceil$ represents here the ceiling function. This limitation on the sum's range is due to the fact that, because of the deltas in the $\omega $-integral, the relevant values of $N$ are the ones for which $\omega_N$ lays in the interior of the $\omega$-integration interval, meaning $\omega_N > m_g$.

\subsection{Non-zero $Q-$components}

In this paragraph, we discuss the expressions of the non-zero components of the quadrupolar Fourier coefficients $Q^{ij}_N$ \cite{peters1963gravitational}. Using equations \eqref{qfourier} and \eqref{qtspace}, we have
\begin{equation}
    Q_N^{ij} = \frac{\mu}{P_b} \int_0 ^{P_b} dt \, e^{- i \omega_N t} r^i (t) \, r^j(t) \,.
\end{equation}
Let's observe that since we chose our frame so that the motion is in the $x-y$ plane, the only non-zero components will be $Q_N^{xx}$, $Q_N^{yy}$ and $Q_N^{xy}=Q_N^{yx}$. For example, the $xx-$component is
\begin{eqnarray}
    Q_N^{xx} &=& \frac{\mu}{P_b} \int_0 ^{P_b} dt \, e^{- i \omega_N t} (r^x (t))^2 \\
    &=& \frac{\mu b^2}{P_b} \int_0 ^{P_b}  dt \, e^{i N \Omega t} (\cos\phi -e)^2 \nonumber \,.
\end{eqnarray}
The main idea to solve those kind of integrals is to first change the integration variable to $\phi$ and then express everything in terms of Bessel functions $J_N (Ne)$. \\
At the end, all the non-zero components will have the same structure
\begin{equation} \label{QforL}
    Q_N^{ij} =\frac{\mu b^2}{2N} L^{ij}_N \,.
\end{equation}
The explicit exprresions of the non-zero $L^{ij}_N$ in functions of the $J_N (Ne)$ are
\begin{eqnarray} \label{Tcomp}
    && L^{xx}_N= J_{N-2} - 2e J_{N-1} +2e J_{N+1} -J_{N+2}  \,, \\
    && L^{yy}_N= J_{N+2} - 2e J_{N+1}-\frac4N J_N +2e J_{N-1}  -J_{N-2} \, , \nonumber \\
    && L^{xy}_N=-i \sqrt{1-e^2} \left (J_{N-2} - 2 J_{N} +J_{N+2} \right )\nonumber  \,,
\end{eqnarray}
where, to find them, we made use of the Bessel function identity $J_N(t) = \frac{1}{2\pi} \int _0 ^{2\pi} e^{i(N \phi -t \sin\phi )}d\phi$, and we denoted with a prime over $J_N$ its derivative respect to its argument, that in this case is always the product $N  e$.\\

\section{Angular Integral $\int d\Omega_k$}

At this point, we are ready to perform the angular integral in \eqref{dedt4}, which at the end will only involve the $\Lambda^{ijkl}$ structure, since, after the consideration we have made on the EMT, it is the only object depending on $\hat k$
\begin{equation}
    \Pi^{ijkl} (\omega_N,\hat n)= \int d\Omega_k \, \Lambda^{ijkl}(\omega_N, \hat k, \hat n) \, .
\end{equation}
We can break down this calculation as the calculation of the integrals of the type
\begin{equation} \label{Iijm}
    I^{i...j} _m= \int d\Omega_k \frac{\hat k^i ... \hat k ^j}{R^m} \,,
\end{equation}
that we encounter when integrating the expression \eqref{TTlambda}. This task can be simplified by exploiting the $SIM(2)$ symmetry of our expressions. \\
Let's make a simple example: we want to calculate the integral $I^i_m$. After the $\hat k-$integration there is only one possible tensorial structure to which the result can be proportional, which is clearly $\hat n^i$. Therefore, we will have
\begin{equation}
        I^{i}_m = B_m \hat n^i \,.
\end{equation}
Contracting with $\hat n^i$, we get the coefficient expression
\begin{equation}
    \hat n^i I^{i}_m = \int d\Omega_k \frac{\hat n^i \hat k^i }{R^m}= I_m^1 =B_m  \,,
\end{equation}
where we defined another type of integrals
\begin{equation}\label{Ipm}
    I^p_m= \int d\Omega_k \frac{(\hat n \cdot \hat k)^p}{R^m} = 2 \pi \int_{-1}^1 dx \frac{x^p}{(1 - \rho\, x)^m} \,.
\end{equation}
In the last expression we used the fact that, since we are integrating over all directions of $\hat k$, the result will not depend on the frame in which we are doing the integral. Thus, we choose the frame such that $\hat n \cdot \hat k = \cos \theta = x$. \\
We include the final explicit expression for the angular integral in Eq.\eqref{angInt}. Note that, one can easily verify it to have the correct GR-limit when sending $m_g \to 0$.

\section{Radiated Power and Period Decrease Rate}

At this point, the ‘‘updated" formula for the gravitational radiated power in VSR is
\begin{eqnarray} \label{dedt5}
    \dot E_{VSR} &=& \frac{G^4 \mu^2 M^3 }{16 \pi b^5} \sum_{N_{min}}\rho(\omega_N) \, N^4 \, L^{*\,ij} L^{ kl} \Pi^{ijkl}|_{\omega = \omega_N} \nonumber \\
    &=& \frac{32 G^4 m_1^2 m_2^2 M }{5 b^5} \sum_{N_{min}} \frac{5 \rho}{512 \pi }  N^4 \, \, L^{*\,ij} L^{ kl} \Pi^{ijkl} \nonumber \\
    &=& \frac{32 G^4 m_1^2 m_2^2 M }{5 b^5} \sum_{N_{min}} f(N ,e, \delta , \hat n) \,, 
\end{eqnarray}
where we defined $\delta \equiv m_g /\Omega$ and the function
\begin{equation}\label{fnn0e}
    f(N, e,\delta ,\hat n) =\frac{5 \, \rho }{512 \pi }N^4 \, \, L^{*\,ij} L^{ kl} \Pi^{ijkl} \, .
\end{equation}
From this formula, we can easily derive the orbital period derivative $\dot P$, since we know that in a keplerian system we must have
\begin{equation}
    \dot P= - 6 \pi \frac{ b^{\frac52 }G^{-\frac32} }{ m_1 m_2 \sqrt{m_1+m_2} } \dot E \, .
\end{equation}
Therefore, after a little manipulation, the rate of period decrease in VSR can be rewritten in the following (experimentally) convenient form
\begin{equation} \label{dpdt}
    \dot P_{VSR} = -\frac{192 \pi \, T_{\odot}^{\frac53}}{5}  \frac{\tilde m_1 \tilde m_2}{\tilde M ^{\frac13}} \left (\frac{P_b}{2\pi} \right )^{-\frac53} \sum_{N_{min}} f(N, e, \delta, \hat n) ,
\end{equation}
with $T_\odot \equiv G M_\odot / c^3 = 4.925490947 \, \mu s $ \cite{weisberg2016relativistic} and the ‘‘tilde"-masses defined as $\tilde m \equiv m / M_\odot$, with $M_\odot$ being the solar mass.

\subsection{Zero-mass limit}

It is easy to see that, when taking the limit $m_g \to 0$, Eq.\eqref{dpdt} reduces to the usual GR formula for the period decrease rate of a keplerian two-body system \cite{peters1963gravitational}
\begin{equation} \label{periodGR}
    \dot P_{GR} = -\frac{192 \pi \, T_{\odot}^{\frac53}}{5}  \frac{\tilde m_1 \tilde m_2}{\tilde M ^{\frac13}} \left (\frac{P_b}{2\pi} \right )^{-\frac53} \frac{1+\frac{73}{24}e^2+\frac{37}{96}e^4}{(1-e^2)^{\frac72}} \,.
\end{equation}
In fact, it can be shown that in this limit the contraction $LL\Pi$ takes exactly the GR standard form, implying 
\begin{equation}
    \lim _{\delta\to 0} f(N,e,\delta,\hat n) = g(N,e) \, ,
\end{equation}
where the expression for $g(N,e)$ is given in Eq.\eqref{gNe}. Therefore, since the following sum-result holds \cite{peters1963gravitational} 
\begin{equation}
    \sum_{N=1}^{+\infty} g(N,e) = \frac{1+\frac{73}{24}e^2+\frac{37}{96}e^4}{(1-e^2)^{\frac72}}\,,
\end{equation}
we end up with the correct GR limit for $m_g\to0$.

\subsection{Case $\hat n // \hat z$} \label{nppz}

\begin{figure}[h!!]
\centering
 \includegraphics[width=8.6cm]{"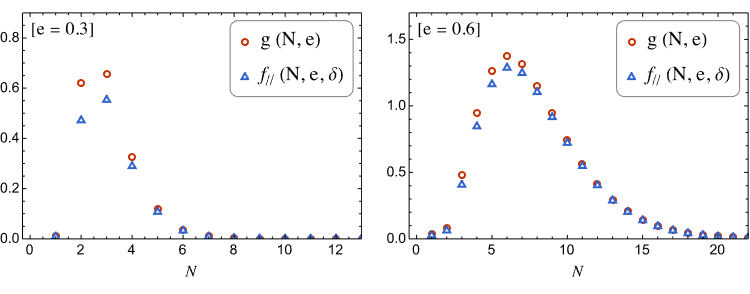"}
    \caption{Comparison between $f_{//}(N,e,\delta)$ and $g(N,e)$ in function of $N$, respectively for the values $e=0.3$ and $e=0.6$, in the exemplificative case of $\delta=0.5$}
    \label{fig:f//vsN}
\end{figure}

Let's start by analyzing analitically the simplest scenario, that is when $\hat n // \hat z$. In this case, all the terms in the $LL \Pi$ contraction proportional to $\hat n$ cancel, since the non-zero $L-$components only involve $x$ and $y$. \\
Thus, we are only left with
\begin{eqnarray}
    && L^{*\,ij}_N L^{ kl}_N \, \Pi^{ijkl} = \\  &&= \frac{2\pi}{15} \left \{ \left (16 \rho^4 + \frac52 \rho^2 -\frac{45}{2} \right. \right. \nonumber \\
    &&\left . \;\;\;\;\;\;\;\;\;\;\;\;\;\;\; + 15 \, \frac{1+2\rho^2-3\rho^4}{2\rho} \tanh^{-1}\rho \right ) L^{*\,ii}_N L^{jj}_N  \nonumber \\ 
    && \;\;\;\;\;\;\;\;\;\;\;\; + 2 \left (16 \rho^4 -\frac{155}{2}\rho^2 +\frac{135}{2} \right . \nonumber\\
    && \;\;\;\left. \left. \;\;\;\;\;\;\;\;\;\;\;\;\;\; -15 \, \frac{7-10\rho^2+3\rho^4}{2\rho} \tanh^{-1}\rho \right ) L^{*\,ij}_N L^{ij}_N \right \} , \nonumber
\end{eqnarray}
while for the contracted $LL-$combinations appearing above we have
\begin{eqnarray}
    L^{*\,ii}_N L^{jj}_N &=& \frac{16}{N^2} J^2_N \, ,\\
    L^{*\,ij}_N L^{ij}_N &=& 2 \left [ (1-e^2)(J_{N-2} -2 J_N +J_{N+2} \,)^2  \right. \nonumber \\
    &&  \;\;\;\;\;\;\;\,+ \frac{4}{N^2} J^2_N +(J_{N-2} -2 e J_{N-1} + \frac2N J_N \nonumber\\
    && \;\;\;\;\;\;\;\;\;\;\;\;\;\;\;\;\;\;\;\;\;\;\;\;\;\;\;\;\;\;\;\;\;\; \left. +2 e J_{N+1}-J_{N+2})^2 \, \right ] ,\nonumber 
\end{eqnarray}
where the argument of all the above Bessel functions $J$ is still $N e$. Note that, being this contribution to $LL\Pi$ independent of $\hat n$, it is also present in the more general case where $\hat n$ is oriented differently. The full expression for $f_{//} \equiv f\,|_{\, \hat n// \hat z}$ is given in Eq.\eqref{f//fperp}. \\
The discrepance respect to GR obtained by considering $f_{//}$ is always negative, resulting in a reduction in the final period decrease rate. This effect can also be appreciated in Fig.\ref{fig:f//vsN}, where we compare the behaviour of $f_{//}$ and $g$ in function of $N$, for $\delta=0.5$ and for two different illustrative eccentricity's values.
\\

\begin{figure}[h!!]
 \includegraphics[width=6.5cm]{"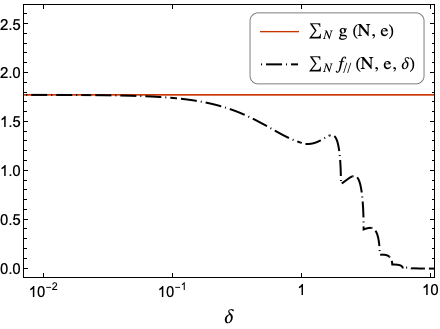"}
    \caption{Behaviour of $\sum_{N_{min}} f_{//}(N,e,\delta)$ and $\sum_{N=0} g(N,e)$ in function of $\delta$, for the illustrative value of $e=0.3$. The horizontal axis is represented in logarithmic scale. It is possible to see that already for $\delta \sim 10$ the high-intensity $N$-values have been excluded from the sum, implying a crucial drop in the total emitted radiation.}
    \label{fig:F//vsD}
\end{figure}

\begin{figure}[h!!]
 \includegraphics[width=8.7cm]{"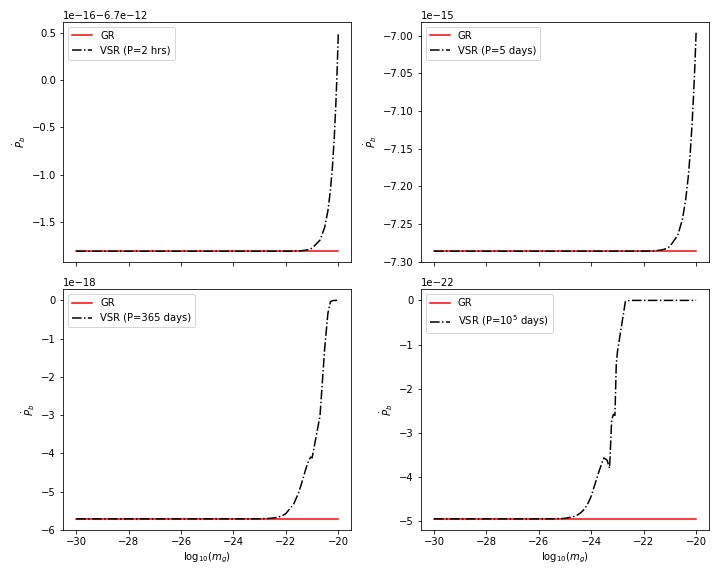"}
    \caption{Period decrease rate in GR and VSR (for $\hat n // \hat z$) in function of $m_g$ and for different values of the orbital period $P_b$. The eccentricity here is set to $e=0.3$.}
    \label{fig:P//vsMg}
\end{figure}

From these graphs, we also recognize the fast-convergency behaviour of $f_{//}$ when increasing $N$. This feature allow us to cut off the $N-$summation (around $ \sim 100$) 
much before the upperlimit given in \eqref{omegaupper}, which for typical binaries' velocities would give $N_{max} \sim 1000$. However, since the $N-$value of the intensity peak of $f_{//}$ increases with $e$, one should be more careful when considering this approximation with very high-eccentricity binaries.\\
There is also another interesting effect coming from the essence of $N_{min}$ in the sum: if $m_g$ is not exactly zero in nature, then, there should exist binary systems for which $\delta>1 $, implying the modes with $N< N_{min}$ would be excluded from the sum, leading to a drop in the emission intensity, as shown in Fig.\ref{fig:F//vsD}, where to each integer value of $\delta$ corresponds a dip in magnitude. Due to the horizontal shift of the intensity peak of $f_{//}$ when increasing $e$, this effect is more easily noticeable for small eccentricities. This fact may lead to the possibility of placing better constraints on the mass $m_g$ from measurements of the period decrease rate for binaries with longer period and smaller eccentricity. See for example Fig.\ref{fig:P//vsMg} where, by taking larger values of the period $P_b$ and therefore increasing $N_{min}$, the effect of progressively excluding the first $N-$values from the sum becomes more and more evident.\\
\\
Note that, even if we have not seen any mention about it in literature, this is a phenomena that should affect not only the VSR realization of linearized massive gravity but other formulations (like \cite{Poddar:2021yjd}) too, since it is simply derived from considerations on the $\omega-$integration range.

\subsection{Generic $\hat n$ case}

\begin{figure}[h!!]
 \includegraphics[width=8.5cm]{"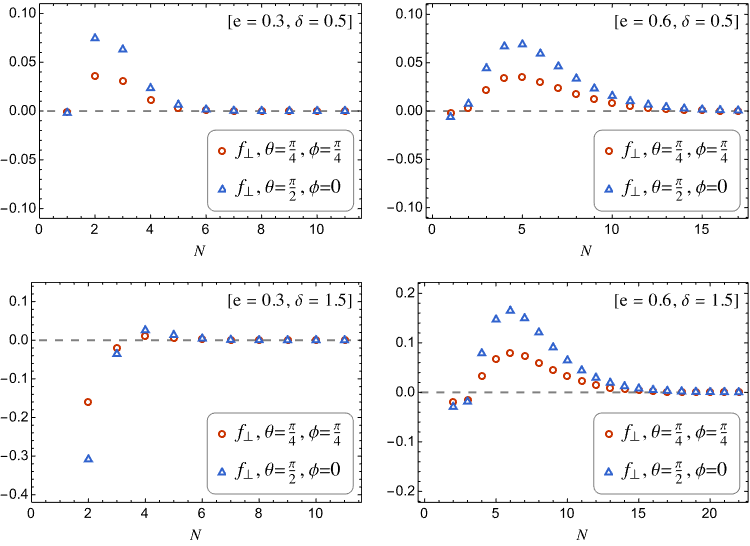"}
    \caption{Behaviour of $f_{\perp}$ in function of $N$, for different values of $\{ e , \delta, \theta , \phi\}$. Here, the qualitative behaviour does not change much by varying $\{\theta,\phi\}$, it just scales the magnitude of the VSR effects. Interestingly, having a greater value of $e$ delays the shift to negative values for the lowest $N-$values when increasing $\delta$.}
    \label{fig:fperpvsN}
\end{figure}

\begin{figure}[h!!]
 \includegraphics[width=6.5cm]{"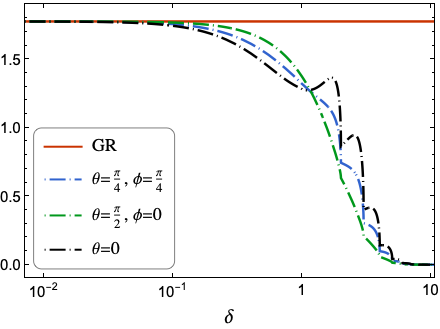"}
    \caption{Comparison of the curves produced by $\sum_{N_{min}} f$ in function of $\delta$, for different values of $\{ \theta , \phi\}$ and with fixed eccentricity $e=0.3$. The horizontal axis is represented in logarithmic scale for convenience.}
    \label{fig:FperpvsD}
\end{figure}

For the generic case is much more difficult to obtain a compact formula. Nevertheless, observe that we have at least a factorization of the VSR effects generated by the generic $\hat n$. In fact, we will have
\begin{equation}
    f(N,\delta,e, \hat n) = f_{//}(N,\delta,e) + f_{\perp}(N,\delta,e ,\hat n) \,,
\end{equation}
where $f_{//}$ is the contribution previously found and $f_{\perp}$ is the new part. The $\hat n-$dependence can be parametrized in function of the two angles $\{\theta,\phi\}$  by choosing
\begin{equation}
    \hat n = (\sin\theta \cos\phi, \sin\theta \sin\phi, \cos\theta)\,,
\end{equation}
with $\theta \in (0,\pi) \;,\; \phi \in (0,2\pi)$. Naturally the case $\hat n // \hat z$ is recovered by taking $\theta = 0$. \\
\\
We include the expression for $f_{\perp}$ in Appendix \ref{genericfcalc}. By plotting $f_{\perp}$ versus $N$ for some indicative values of $\{ \delta , e , \theta , \phi\}$, we notice that it is mostly close to zero or positive as long as $\delta$ is small, while fot the first few available $N-$values it undergoes a progressive shift to negative values when increasing $\delta$. This feature is shown, for example, in Fig.\ref{fig:fperpvsN}. Thus, in contrast to the VSR contribution $f_{//}-g$, which was always negative, $f_{\perp}$ does not have in general a definite sign.\\
\\
Another interesting fact that can be graphically appreciated is that, starting from small values of $\delta$, the $f-$contribution in the case $\hat n//\hat z$ is always below its more generic counterpart with $\hat n$ oriented differently. This situation changes when increasing $\delta$, in which case the hierarchy gets turned over, as we can see in Fig.\ref{fig:FperpvsD}.

\section{Graviton mass bounds from data}\label{sec:mgbound}

\begin{figure}[h!!]
    \includegraphics[width=8cm]{"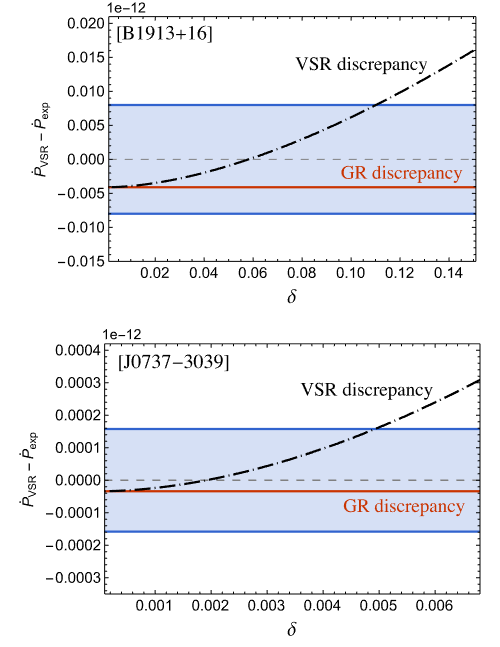"}
    \caption{Discrepancy in the period decrease rate predicted by GR and VSR calculations respect to the experimental values measured for the Hulse-Taylor and the Double Pulsar. The light-blue band here represent the experimentally allowed discrepancy region at $95\%$ confidence level.}
    \label{fig:massbounds}
\end{figure}

In this section, we want to use the previous results to show how they allow to bound experimentally the graviton mass parameter in VSR. \\
For this purpose, we will use data from two of the most well-studied pulsar binaries: the Hulse-Taylor binary PSR B1913+16, the first binary pulsar ever discovered \cite{hulse1975discovery}, and the Double Pulsar PSR J0737-3039A/B. We include information on these two systems in Table \ref{tab:pulsarsdata}.
\begin{table}[ht]
\centering
\begin{tabular}[t]{lcc}
\toprule
{Pulsar} & {B1913+16} \cite{weisberg2016relativistic} & {J0737-3039} \cite{kramer2021strong} \\ \midrule
    {$m_1 (m_\odot)$} & 1.438(1) & 1.33819(1)  \\
    {$m_2 (m_\odot)$}  & 1.390(1)  & 1.24887(1)  \\
    {$P_b (d)$} & 0.322997448918(3) & 0.102251559297(1) \\
    {$e$} & 0.6171340(4)  & 0.08777702(6) \\ \midrule
    {$\dot P_{exp}  (s s^{-1}) \times 10^{12}$}  & {$-2.398$}   & {$-1.247782$}  \\
    {$\sigma_{\dot P}  (s s^{-1}) \times 10^{12}$}  & 0.004  & 0.000079  \\
\bottomrule
\end{tabular} 
\caption{Descriptive data for the two pulsar binaries selected for this work. Parenthesized numbers represent the $1\sigma$ uncertainty in the last digit quoted. The sources of the data for each pulsar are cited alongside their names.}
\label{tab:pulsarsdata}
\end{table} \\
A more precise inspection of the problem would require a statistical in depth analysis of the general VSR formula on a dataset including a much wider collection of binaries, but that is not the purpose of this paper.\\
The fundamental frequency $\Omega$ of both binaries taken into consideration when translated in terms of energies is approximately $ \sim 10^{-19} eV$. We will, then, suppose to be in the small $\delta \equiv m_g / \Omega $ regime. This assumption is reasonable since the kinematical bound $m_g \lesssim 10^{-22} eV$ \cite{baker2017strong} obtained from the combined observation of the events GW170817 and GRB 170817A should hold also in VSR: in fact, the on-shell graviton's dispersion relation in gauge-fixed VSLG take the usual SR structure \cite{grav3} independently of the $\hat n- $direction.\\ 
For the considered binaries, we have seen that, when $\delta$ is small, the period decrease rate in the $\hat n // \hat z-$scenario is always bigger (in absolute value) than the one for the other $\hat n -$orientations. Thus, we restrict our analysis to $\hat n // \hat z$, since it implies the greatest VSR discrepancy possible in our context, and therefore leads to the stringest bound on graviton mass obtainable taking VSR for granted.\\
That being the case, the procedure used for the estimation of the $m_g- $upperlimit is simple: for small $\delta$, the  $\hat n // \hat z- $contribution gets bigger in absolute value with $\delta$, as shown in Fig. \ref{fig:F//vsD}. Then, it will be sufficient to increase the value of $m_g $ to the point where we saturate the maximal discrepancy allowed by a $95\%$ confidence interval around the experimental rate of period decrease $\dot P_{exp}$ due to GWs emission. The value found through this process will represent our constraint. See Fig.\ref{fig:massbounds} for a graphical representation. \\
Let's start from the pulsar PSR B1913+16. In this case, following the above procedure we obtain a $\delta-$upperlimit of $\delta < 0.1102 $, which corresponds to the following bound on the graviton mass
\begin{equation}
    m_g \lesssim 5.2 \times 10^{-20} eV \,.
\end{equation}
For the other binary under consideration, PSR J0737-3039A, the upperbound that we get is
\begin{equation}
    m_g \lesssim 2.3 \times 10^{-21} \, ,
\end{equation}
which, while being more stringent than the first one, is still not small enough to be an improvement respect to the kinematical bound obtained from the above-mentioned detection of GW170817.\\
 \\
Nevertheless, it would be straightforward to improve these constraints by either further enhancing the precision of the measurements of the period decrease rate or by focusing the attention to binaries with a large orbital size and therefore longer orbital periods. In fact, the longer the period, the bigger the $\delta-$ratio at fixed $m_g$, leading in principle to larger VSLG effects. Furthermore, as already mentioned in subsection \ref{nppz}, small values of the eccentricity could also be beneficial for these tests since the high-intensity modes in the $\sum_N$ would be excluded for smaller values of $m_g$, keeping $P_b$ fixed. In this way, one could hope to detect some novel VSR effect or, at least, place better constraints on the VSR origin of a graviton mass.

\section{Summary and Conclusions}

In this paper, using an effective field theory approach for VSLG, we derived expressions for the rates of energy loss and orbital period decrease for a binary system. \\
While in the case of many other massive gravity models the massless limit does not give the GR result, we have observed that in VSLG we easily recover the standard GR formula in the zero-mass limit. This result was expected since, in contrast to what happens for example with the Fierz Pauli formulation, in VSR we do not introduce additional d.o.f and therefore we have no new polarizations which can contribute to the energy loss when sending $m_g \to 0$. In this sense, VSLG would represent an “healthy" massive graviton alternative.\\
Afterwards, we proceeded to study and analyze some peculiar feature of the new VSR formula for the period decrease rate, both in the simplest case of $\hat n // \hat z$ and the more generic case of an $\hat n$ oriented differently. Interestingly, we also higlighted a “dumping" effect that we think should be present also in other massive gravity formulations, since it is generated simply by the relation among the fundamental frequency $\Omega$ and the graviton mass $m_g$.\\
Finally, we applied our results on some real binary systems to compare them with available experimental data and put bounds on the VSR graviton mass: for the two binary stars took into consideration, the Hulse-Taylor binary and PSR J0737-3039, we found an upperbound of about $\sim 10^{-21 }eV$, which would be almost as strong as the kinematical bound from GW170817 \cite{baker2017strong}.
The values we obtained are comparable to other graviton mass upperlimits from binary systems found in literature, like \cite{Poddar:2021yjd, finn2002bounding, miao2019bounding}, since usually the most relevant radiation contribution is of quadupolar nature. For some particular models, like the one in \cite{shao2020new}, much stronger bounds are obtained due to the additional presence of dipolar modes. Furthermore, we note that the constraints obtained here for a VSR origin of graviton mass could be improved in the future not only by increasing experimental precision in the period decrease rate measurements or studying binaries with longer periods, but for example also by exploiting direct GW detections, for which further study in the VSLG framework should be carried on.

\acknowledgments
A.Santoni thanks T.K.Poddar for fruitful discussion at ICHEP 2022 and acknowledges financial support from ANID Fellowship CONICYT-PFCHA/DoctoradoNacional/2020-21201387. J.A. acknowledges the partial support of the Institute of Physics PUC and Fondo Gemini Astro20-0038.

\appendix

\section{VSLG Lagrangian and Feynman Rules} \label{appLagandFeyn}

In this appendix we include some useful details and explicit expressions of VSLG \cite{grav3}. The full lagrangian $\mathcal{L}$ considered in our approach 
\begin{eqnarray}
\label{actionvsr}
 \mathcal{L} &=& \frac{1}{2} h^{\mu \nu} \mathcal{O}_{\mu \nu \alpha \beta}
h^{\alpha \beta} - \frac{\kappa}{2}h_{\mu \nu} T^{\mu \nu} \\
&& + \xi \partial_{\mu} \left( h^{\mu \nu} - \frac{1}{2}
\eta^{\mu \nu} h \right) \partial^{\lambda} \left( h_{\lambda \nu} -
\frac{1}{2} \eta_{\lambda \nu} h \right) \,,\nonumber
\end{eqnarray}
is constructed with the quadratic operator $\mathcal{O}$ for the graviton \eqref{hlag}, the matter coupling \eqref{lagmatt} and the gauge fixing term \eqref{gaugefixing}. Up to a rescaling constant which can be absorbed in the $h-$field, the full expression for the operator $\mathcal{O}$ obtained in VSLG is given by
\begin{widetext}
\begin{eqnarray} \label{operatorO}
\mathcal{O}_{\mu \nu \alpha \beta} &=& - \partial_{\mu} \partial_{\nu} \eta_{\alpha
\beta} + \frac{1}{2} \partial_{\mu} \partial_{\alpha} \eta_{\nu \beta} +
\frac{1}{2} \partial_{\mu} \partial_{\beta} \eta_{\nu \alpha} -
\partial_{\alpha} \partial_{\beta} \eta_{\mu \nu} + \frac{1}{2} \partial_{\nu}
\partial_{\beta} \eta_{\mu \alpha} + \frac{1}{2} \partial_{\nu}
\partial_{\alpha} \eta_{\mu \beta} + \partial^2 \eta_{\mu \nu} \eta_{\alpha
\beta} - \frac{1}{2} \partial^2 \eta_{\mu \alpha} \eta_{\nu \beta} \nonumber\\
& & - \frac{1}{2} \partial^2 \eta_{\mu \beta} \eta_{\nu \alpha} + m^2_g \eta_{\mu
\nu} \eta_{\alpha \beta} - \frac{m^2_g}{2} \eta_{\mu \alpha} \eta_{\nu \beta}
- \frac{m^2_g}{2} \eta_{\mu \beta} \eta_{\nu \alpha} - m^2_g N_{\mu} N_{\nu}
\partial_{\alpha} \partial_{\beta} + \frac{m^2_g}{2} N_{\mu} N_{\alpha}
\partial_{\nu} \partial_{\beta} + \frac{m^2_g}{2} N_{\mu} N_{\beta}
\partial_{\nu} \partial_{\alpha} \nonumber\\
& & + \frac{m^2_g}{2} N_{\nu} N_{\alpha}
\partial_{\mu} \partial_{\beta} + \frac{m^2_g}{2} N_{\nu} N_{\beta}
\partial_{\mu} \partial_{\alpha} - m^2_g N_{\alpha} N_{\beta} \partial_{\mu}
\partial_{\nu} + m^2_g \partial^2 N_{\mu} N_{\nu} \eta_{\alpha \beta} -
\frac{m^2_g}{2} \partial^2 N_{\mu} N_{\alpha} \eta_{\nu \beta} -
\frac{m^2_g}{2} \partial^2 N_{\mu} N_{\beta} \eta_{\nu \alpha} \nonumber\\
& & - \frac{m^2_g}{2} \partial^2 N_{\nu} N_{\beta} \eta_{\mu \alpha} -
\frac{m^2_g}{2} \partial^2 \eta_{\mu \beta} N_{\nu} N_{\alpha} + m^2_g
\partial^2 N_{\alpha} N_{\beta} \eta_{\mu \nu} - m^2_g \eta_{\mu \nu}
N_{\alpha} \partial_{\beta} - m^2_g \eta_{\mu \nu} \partial_{\alpha} N_{\beta}
+ \frac{m^2_g}{2} \eta_{\mu \alpha} N_{\nu} \partial_{\beta} \nonumber\\
& & + \frac{m^2_g}{2}
\eta_{\mu \alpha} \partial_{\nu} N_{\beta} + \frac{m^2_g}{2} \eta_{\mu \beta}
N_{\nu} \partial_{\alpha} + \frac{m^2_g}{2} \eta_{\mu \beta} \partial_{\nu}
N_{\alpha} + \frac{m^2_g}{2} \eta_{\nu \alpha} N_{\mu} \partial_{\beta} +
\frac{m^2_g}{2} \eta_{\nu \alpha} \partial_{\mu} N_{\beta} + \frac{m^2_g}{2}
\eta_{\nu \beta} N_{\mu} \partial_{\alpha} \nonumber\\
& & + \frac{m^2_g}{2} \eta_{\nu \beta}
\partial_{\mu} N_{\alpha} - m^2_g \eta_{\alpha \beta} N_{\mu} \partial_{\nu} -
m^2_g \eta_{\alpha \beta} \partial_{\mu} N_{\nu} \, ,
\end{eqnarray}
\end{widetext}
which corresponds to the rescaled position-space version of the formula $(4)$ for the momentum-space operator $\mathcal{O}$ in \cite{grav3}.\\
From the lagrangian \eqref{actionvsr} and the EMT given by \eqref{emtscalar}, the momentum-space Feynman rules for the propagator and vertex $h\phi \phi$ respectively are
\begin{widetext}
\begin{eqnarray}
i\mathcal{O}_{\rho \sigma \alpha \beta}^{- 1} &=& \frac{i P_{\rho \sigma \alpha
\beta}}{k^2 - m^2_g} + \frac{i m^4_g}{2 \, k^4 (k^2 - m^2_g)} \Lambda_{\rho \sigma} \Lambda_{\alpha
\beta} \nonumber\\
& & - \frac{i m^2_g}{2 \, k^2 (k^2 - m^2_g)} \bigg(\Lambda_{\rho \alpha} \eta_{\sigma
\beta} + \Lambda_{\rho \beta} \eta_{\sigma \alpha} + \Lambda_{\sigma \alpha}
\eta_{\rho \beta} + \Lambda_{\sigma \beta} \eta_{\rho \alpha} - \Lambda_{\rho
\sigma} \eta_{\alpha \beta} - \Lambda_{\alpha \beta} \eta_{\rho \sigma} \bigg)\nonumber\\
& & + \frac{i m^2_g}{k^2 (k^2 - m^2_g)} \bigg[ - \frac{1}{2} (N_{\rho} k_{\sigma}
+ N_{\sigma} k_{\rho})(N_{\beta} k_{\alpha} + N_{\alpha} k_{\beta}) +
N_{\rho} N_{\sigma} k_{\alpha} k_{\beta} + N_{\alpha} N_{\beta} k_{\rho}
k_{\sigma} \bigg] \, , \label{propagator}
\end{eqnarray}
\end{widetext}
and
\begin{equation}
V_{h \phi \phi} = - i \frac{\kappa}{2} (k^{\rho} p^{\sigma} + k^{\sigma}
p^{\rho} - \eta^{\rho \sigma} (k \cdot p - m^2)) \, .
\end{equation}
Here, we defined
\begin{eqnarray}
P_{\rho \sigma \alpha \beta} &=& \frac{1}{2} (\eta_{\rho \alpha} \eta_{\sigma
\beta} + \eta_{\rho \beta} \eta_{\sigma \alpha} - \eta_{\rho \sigma}
\eta_{\alpha \beta}) \, ,\\
\Lambda_{\rho \sigma} &=& k_{\rho} N_{\sigma} +
k_{\sigma} N_{\rho} - N_{\rho} N_{\sigma} k^2 \, .
\end{eqnarray}
We observe that we recover the standard propagator in the limit $m_g\to 0$ \cite{Donoghue:2017pgk}.

\section{Polarizations' sum in VSLG} \label{appPolar}

Here we include some details on the calculation of the polarization sum $S^{\mu\nu\alpha\beta}$ in VSLG. For this calculation, we exploit the covariance of $S$ to determine the allowed structures: the available tensorial “building blocks" from which we start are the flat metric $\eta$, the four-momentum $k$ and the four-vector $N^\mu \equiv n^\mu / n \cdot k$ as usual in VSR. Simbolically, we obtain
\begin{eqnarray}
    S &=& 3\, \eta\eta  + 6\, \eta \, k k + + 6\, \eta N N + 6\, k k N N + 12\, \eta \, k \,N \nonumber \\
    &&+ 4\, k k k N + 4\, k N N N  + k k k k + N N N N\, ,
\end{eqnarray}
where the numbers represent the different indices combinations for each possible tensorial structure.\\ 
At this point, from the gauge conditions already imposed in VSLG \cite{grav3}: $k_\mu h^{\mu\nu}= n_\mu h^{\mu \nu}=0$ and $h=h^\mu_{\,\mu}=0$, we deduce the following constraints on $S$
\begin{equation}
\begin{cases}
k_\mu S^{\mu\nu\alpha\beta} = 0 \,, \\
n_\mu S^{\mu\nu\alpha\beta} = 0 \,, \\
\eta_{\mu\nu} S^{\mu\nu\alpha\beta} = 0 \, .
\end{cases} 
\end{equation}

Pairing this conditions with the one deriving from normalization $S^{\mu \nu}_{\;\;\;\;\mu \nu}= \sum_\lambda \epsilon_\lambda ^{\,\mu\nu} \epsilon^*_{\lambda\, \mu \nu} =2$, which ensures the correct limit for $m_g\to 0$, and the index simmetries $\mu \Longleftrightarrow \nu$, $\alpha \Longleftrightarrow \beta$, $\mu \nu \Longleftrightarrow \alpha \beta$ (since $S$ is related also to the $h-$propagator \cite{Poddar:2021yjd}), we finally find the expression
\begin{widetext}
\begin{eqnarray}
 S^{\mu \nu \alpha \beta} &=& \frac12 g^{\mu \alpha} g^{\nu \beta}+ \frac12 g^{\mu \beta} g^{\nu \alpha} -\frac12 g^{\mu\nu} g^{\alpha\beta} \\
 && +\frac{m^2_g}{2} g^{\nu\beta} N^\mu N^\alpha +\frac{m^2_g}{2} g^{\mu\beta} N^\nu N^\alpha +\frac{m^2_g}{2} g^{\nu \alpha} N^\mu N^\beta + \frac{m^2_g}{2} g^{\mu\alpha} N^\nu N^\beta - \frac{m^2_g}{2} g^{\mu\nu} N^\alpha N^\beta -\frac{m^2_g}{2} g^{\alpha\beta} N^\mu N^\nu \nonumber \\
 && +\frac12 g^{\alpha\beta} N^\nu k^\mu -\frac12 g^{\nu\beta} N^\alpha k^\mu -\frac12 g^{\nu\alpha} N^\beta k^\mu + \frac12 g^{\alpha\beta} N^\mu k^\nu -\frac12 g^{\mu\beta} N^\alpha k^\nu -\frac12 g^{\mu\alpha} N^\beta k^\nu \nonumber \\
 &&-\frac12 g^{\nu\beta} N^\mu k^\alpha -\frac12 g^{\mu\beta} N^\nu k^\alpha + \frac12 g^{\mu\nu} N^\beta k^\alpha -\frac12 g^{\nu \alpha} N^\mu k^\beta - \frac12 g^{\mu\alpha} N^\nu k^\beta +\frac12 g^{\mu\nu} N^\alpha k^\beta \nonumber \\
 && +N^\mu N^\nu k^\alpha k^\beta+k^\mu k^\nu N^\alpha N^\beta + \frac{m_g^4}{2} N^\mu N^\nu N^\alpha N^\beta \nonumber \\
 && -\frac{m_g^2}{2} N^\mu N^\nu N^\alpha k^\beta -\frac{m_g^2}{2} N^\mu N^\nu N^\beta k^\alpha -\frac{m_g^2}{2} N^\mu N^\alpha N^\beta k^\nu -\frac{m_g^2}{2} N^\nu N^\alpha N^\beta k^\mu \nonumber \,,
\end{eqnarray}
\end{widetext}
where we also had to consider the on-shell condition $k^2=m^2_g$ for the graviton polarizations' tensors.\\ 
Now, since all indices of $S^{\mu\nu\alpha\beta}$ are contracted with EMTs, the terms proportional to $k^\mu$ gets cancelled due to energy momentum conservation $k^\mu T_{\mu\nu} = 0$. Thus, including only relevant terms, the expression of $S$ becomes
\begin{eqnarray}\label{polsum} 
 S^{\mu \nu \alpha \beta} &=& \frac12 g^{\mu \alpha} g^{\nu \beta}+ \frac12 g^{\mu \beta} g^{\nu \alpha} -\frac12 g^{\mu\nu} g^{\alpha\beta} \\
 &&+ \frac{m^2_g}{2} (g^{\nu\beta} N^\mu N^\alpha + g^{\mu\beta} N^\nu N^\alpha + g^{\nu \alpha} N^\mu N^\beta )  \nonumber\\
 &&+ \frac{m^2_g}{2} (g^{\mu\alpha} N^\nu N^\beta - g^{\mu\nu} N^\alpha N^\beta - g^{\alpha\beta} N^\mu N^\nu ) \nonumber \\
 && + \frac{m_g^4}{2} N^\mu N^\nu N^\alpha N^\beta \nonumber \,. 
\end{eqnarray}\\

\section{Expression of $\Lambda^{ijkl} (\omega, \hat k ,\hat n)$} \label{appLamb}

The calculation of $\Lambda^{ijkl}$ starting from $S^{\mu\nu\alpha\beta}$ is quite straightforward: as already stated in the main text, we just have to expand the contraction of $S$ with the EMTs, separe time and spatial parts, and finally transform time indices into spatial ones through \eqref{timetospatial}. \\
Let's give an example by realizing this calculation for the term $\sim \tilde T^2$, with $\tilde T = \tilde T^\mu_{\;\,\mu}$
\begin{eqnarray}
    \tilde T^* \tilde T &=& (\tilde T^{*\, 00} -\tilde T^{*\,ii})(\tilde T^{00}-\tilde T^{jj})\\
    &=& \tilde T^{*\,00} \tilde T^{00} - \tilde T^{*\,00} \tilde T^{ii} - \tilde T^{00} \tilde T^{*\,ii} + \tilde T^{*\,ii} \tilde T^{jj} \nonumber \\
    &=& (\rho^4 \hat k^i \hat k^j \hat k^k \hat k^l -  \rho^2 \hat k^i \hat k^j \delta^{kl} \nonumber \\
    && \;\;\; -\rho^2 \hat k^k \hat k^l \delta^{ij} + \delta^{ij} \delta^{kl}) \tilde T^{*\, i j} \tilde T^{kl} \nonumber \,,
\end{eqnarray}
with $\delta^{ij}= diag \{ 1,1,1\}$ being the identity tensor in three spatial dimensions. The computation of the other terms involved in $\Lambda^{ijkl}$ is analogous. After all the calculations of this kind, we find the following complete expression
\begin{widetext}
\begin{eqnarray} \label{TTlambda}
    \Lambda^{ijkl}(\omega, \hat k ,\hat n) &=& \frac12 \rho^4 \hat k^i \hat k^j \hat k^k \hat k^l-  \rho ^2\hat k^j \hat k^k \delta^{il} -\rho ^2\hat k^l \hat k^i \delta^{kj}  + \frac12 \rho^2 \hat k^i \hat k^j \delta^{kl} + \frac12 \rho^2 \hat k^k \hat k^l \delta^{ij} +\delta^{ik}\delta^{jl}-\frac12 \delta^{ij}\delta^{kl} + \\
    && +\frac{m^2_g}{\omega^2} \frac{1}{R^2}\left [ \rho^4 \hat k^i \hat k^j \hat k^k \hat k^l - 2 \rho^2 \hat k^i\hat k^k \delta^{jl} - \rho^{3} \hat n^j \hat k^i\hat k^k\hat k^l- \rho^{3} \hat n^l \hat k^k\hat k^i\hat k^j +2 \rho \, \hat n^j \hat k^l \delta^{ik}+2 \rho \, \hat n^l \hat k^j \delta^{ik}+2 \rho^2 \hat n^i \hat n^k \hat k^j\hat k^l \right. \nonumber\\
    && \;\;\;\;\;\;\;\;\;\;\;\;\;\;\;\;- 2 \hat n^i \hat n^k \delta^{jl} +\frac12 \rho^2 \delta^{ij } \hat k^k \hat k^l+\frac12 \rho^2 \delta^{kl } \hat k^i \hat k^j - \rho\, \hat n^i \hat k^j \delta^{ k l}- \rho\, \hat n^k \hat k^l \delta^{ ij} \nonumber \\
    &&\left.\;\;\;\;\;\;\;\;\;\;\;\;\;\;\;\; - \frac12 \rho^2 \hat n^i \hat n^j \hat k^k \hat k^l - \frac12 \rho^2 \hat n^k \hat n^l \hat k^i \hat k^j + \frac12 \hat n^i \hat n^j \delta^{kl}+ \frac12 \hat n^k \hat n^l \delta^{ij} \right] \nonumber\\
    && +\frac{m^4_g}{\omega^4} \frac{1}{R^4}\left [\frac12 \rho^4 \hat k^i \hat k^j \hat k^k \hat k^l  + 2 \rho^2 \hat n^i \hat n^k \hat k^j \hat k^l +\frac12 \hat n^i \hat n^j \hat n^k \hat n^l - \rho^{3} \hat n^l \hat k^i \hat k^j \hat k^k - \rho^{3} \hat n^j \hat k^k \hat k^l \hat k^i \right. \nonumber\\
    && \left. \;\;\;\;\;\;\;\;\;\;\;\;\;\;\;\;+ \frac12 \rho^2 \hat n^i \hat n^j \hat k^k \hat k^l + \frac12 \rho^2 \hat n^k \hat n^l \hat k^i \hat k^j -  \rho\, \hat n^i \hat n^j \hat n^k \hat k^l -  \rho\, \hat n^k \hat n^l \hat n^i \hat k^j \right ] \,, \nonumber
\end{eqnarray}
\end{widetext}
which is symmetric under $i j \Longleftrightarrow k l$. In principle, the result could be not symmetric in $i j \Longleftrightarrow k l$, since the indices of $ S$ are contracted with two “different" objects. Anyway, due to the symmetries of $S$, the contraction has the effect of considering only the symmetryzed $TT-$product: $\tilde T^{*\,\mu\nu}\tilde T^{\alpha\beta}\to \frac12(\tilde T^{*\,\mu\nu}\tilde T^{\alpha\beta}+\tilde T^{*\,\alpha\beta}\tilde T^{\mu\nu})$, implying that $\Lambda$ inherit the same simmetries of $S$, among which the pair-interchange symmetry $i j \Longleftrightarrow k l$.\\
Note also that, even if not explicit in expression \eqref{TTlambda}, $\Lambda$ can be symmetrized also respect to $i \Longleftrightarrow j$ and $k \Longleftrightarrow l$, since it is contracted with the spatial components $\tilde T^{i j}$ of the symmetric EMTs.

\subsection{Total Angular Integral $\Pi^{ijkl}(\omega,\hat n)$ } \label{appIijm}
In this section we include the explicit expression of the angular integral $\Pi$ found after calculating, with the help of Mathematica, all the needed integrals of the type $I^{i...j}_n$
\begin{widetext}
\begin{eqnarray} \label{angInt}
\Pi^{ijkl}(\omega,\hat n) &=& \frac{2 \pi}{15} \left(\rho^4+10 \rho^2-15\right) \delta^{ij} \delta^{kl}+\frac{4 \pi}{15} \left(\rho^4-10 \rho^2+15\right) \delta^{ik} \delta^{jl}\\
    &+& \pi (1-\rho^2) \left [-{ \left( 2 \rho^2+1-\frac{3 \rho^2+1}{\rho} \tanh ^{-1}\rho \right)} \delta^{ij}\delta^{kl} \right.  -2 \left( 2 \rho^2-7+\frac{7-3 \rho^2}{\rho} \tanh^{-1}\rho \right) \delta^{ik}\delta^{jl} \nonumber\\
    &+& \frac{1 }{3} \left( 16 \rho^2-3+\frac{9-21 \rho^2}{\rho} \tanh ^{-1}\rho \right) (\hat n^i \hat n^j \delta^{kl} + \hat n^k \hat n^l \delta^{ij}) +\frac{4}{3} \left( 16 \rho^2-39 +\frac{33-21 \rho^2}{\rho} \tanh ^{-1}\rho \right) \hat n^i \hat n^k \delta^{jl} \nonumber\\
    &-& \left. \frac{ 1}{3}\left( 80 \rho^2-117+\frac{111-99 \rho^2}{\rho} \tanh ^{-1}\rho \right)  \hat n^i \hat n^j \hat n^k \hat n^l \right ] \,, \nonumber
\end{eqnarray}
\end{widetext}
which one can check to have the correct limit from linearized GR when sending $\rho \to 1$ or equivalently $m_g\to 0$.
\begin{equation}
    \lim_{m_g\to0} \Pi^{ijkl} = -\frac{8 \pi}{15} \delta^{ij} \delta^{kl}+\frac{8 \pi}{5}  \delta^{ik} \delta^{jl}\, .
\end{equation}


\section{Expressions of $f(N,\delta,e,\hat n)$} \label{genericfcalc}

We use this appendix to collect some details on the final expression assumed by the function $f(N,\delta,e,\hat n)$ and its components. We remind, in fact, that we generally have that $f$ is given by two contributions: $f_{//}$ and $f_{\perp}$
\begin{equation}
    f(N,e,\delta ,\hat n)=f_{//}(N,\delta, e )+ f_\perp(N,\delta, e ,\hat n) \, .
\end{equation}\\
While $f_{//}$ is always present, being independent of $\hat n$, the $f_{\perp}$-part is relevant only when $\hat n$ is not parallel to the $z-$direction. This fact directly implies that $f_{//}$ should reduce to the GR contribution in the limit of $m_g\to0$, while the one coming from $f_{\perp}$ should just vanish. \\
Due to what we said, the simplest expression for $f$ is clearly obtained in the case where $\hat n$ is orthogonal to the orbital plane. In the following, we include the explicit expression for both contributions
\begin{widetext}
\begin{eqnarray} \label{f//fperp}
    f_{//}(N,e,\delta) &=& \sqrt{1-\frac{\delta^2}{N^2}} \left \{  \left [ 1 + \frac{\delta^2}{6 N^2} \left (\frac{91}{2} +\frac{16 \delta^2}{ N^2} -15 \frac{4 + \frac{3\delta^2}{N^2}}{2\sqrt{1-\frac{\delta^2}{N^2}}} \tanh^{-1}\sqrt{1-\frac{\delta^2}{N^2}} \right ) \right ] g(N,e) \right . \\
    && \;\;\;\;\;\;\;\;\;\;\;\;\;\;\;\;\;\;\;\;\;\;\;\;\;\;\;\;\;\; \left. + \frac{ \delta^2 }{48} \left (-\frac{25}{6} +\frac{80 \delta^2}{3 N^2} + 5 \frac{ 4 - \frac{15\delta^2}{N^2}}{2\sqrt{1-\frac{\delta^2}{N^2}}} \tanh^{-1}\sqrt{1-\frac{\delta^2}{N^2}}  \right ) J^2_N (Ne) \right \} \nonumber \,,\\
f_{\perp}(N, e, \delta , \hat n)&=& - \frac{5 N^2}{384} \delta ^2 \sqrt{1-\frac{\delta^2}{N^2}} \left\{  
\left (23 +\frac{16 \delta^2}{ N^2} - \frac{12 + \frac{21\delta^2}{N^2}}{\sqrt{1-\frac{\delta^2}{N^2}}} \tanh^{-1}\sqrt{1-\frac{\delta^2}{N^2}} \right ) \hat n^i \hat n^k L^{*\,ij}_N L^{kj}_N
 \right. \nonumber \\ 
&& \;\;\;\;\;\;\;\;\;\;\;\;\;\;\;\;\;\;
-\left (\frac{37}{4} +\frac{20 \delta^2}{ N^2} - \frac{12 + \frac{99\delta^2}{ N^2}}{4 \sqrt{1-\frac{\delta^2}{N^2}}} \tanh^{-1}\sqrt{1-\frac{\delta^2}{N^2}} \right )n^i \hat n^j \hat n^k \hat n^l L^{*\,ij}_N L^{kl}_N
 \nonumber\\
&& \;\;\;\;\;\;\;\;\;\;\;\;\;\;\;\;\;\;\left.
+\frac{2}{N} \left (13 -\frac{16 \delta^2}{ N^2} - \frac{12 - \frac{21\delta^2}{N^2}}{\sqrt{1-\frac{\delta^2}{N^2}}} \tanh^{-1}\sqrt{1-\frac{\delta^2}{N^2}} \right ) \sin^2\theta (\cos^2\phi \, L^{xx}_N + \sin^2 \phi \, L^{yy}_N) J_N 
 \right \} \, , \nonumber
    \end{eqnarray}
with $g(N,e)$ being the same function defined in \cite{peters1963gravitational} 
\begin{eqnarray} \label{gNe}
    g(N,e) &=& \frac{N^4}{32} \left \{ (J_{N-2} -2 e J_{N-1} + \frac2N J_N +2 e J_{N+1}-J_{N+2})^2 + (1-e^2)(J_{N-2} -2 J_N +J_{N+2} \,)^2 +\frac{4}{3N^2} J^2_N \right \} \,, \nonumber \\
    \end{eqnarray}
\end{widetext}
being $N e$ the argument of the above $J-$functions, and
\begin{eqnarray}
    \hat n^i \hat n^j L^{*\,ik}_N L^{jk}_N &=& \sin^2\theta (\cos^2\phi \, (L^{xx}_N)^2 + \sin^2 \phi \,(L^{yy}_N)^2) \nonumber \\
    && \;\;\;+ \sin^2\theta |L^{xy}_N|^2 \,, \nonumber\\
    n^i \hat n^j \hat n^k \hat n^l L^{*\,ij}_N L^{kl}_N &=& \sin^4\theta (\cos^2\phi \, L^{xx}_N + \sin^2 \phi\, L^{yy}_N)^2\nonumber \\
    && \;\;\;+4 \sin^4 \theta \sin^2\phi \cos^2\phi |L^{xy}_N|^2 \,. \nonumber \\
\end{eqnarray}


\bibliography{apssamp}

\end{document}